\documentclass[aps,pra,showkeys,floatfix,12pt,tightenlines]{revtex4-1}
\usepackage[utf8]{inputenc}
\usepackage[T1]{fontenc}
\usepackage{graphicx}%
\usepackage{silence}
\usepackage{hyperref}
\begin{document}

\title{StrAPS: Structural Angular Power Spectrum for Discovering Novel Morphologies in Block Copolymers}

\author{Dominic M. Robe}
\affiliation{Soft Matter Informatics Research Group, Department of Mechanical Engineering, Faculty of Engineering and Information Technology, The University of Melbourne, Parkville, VIC 3010, Australia}
\author{Elnaz Hajizadeh}
\email{ellie.hajizadeh@unimelb.edu.au}
\affiliation{Soft Matter Informatics Research Group, Department of Mechanical Engineering, Faculty of Engineering and Information Technology, The University of Melbourne, Parkville, VIC 3010, Australia}

\begin{abstract}
The morphologies of phase separating systems have formal distinctions such as symmetry groups, but the analysis protocol for labeling a particular phase field with a morphology requires manual expertise, arbitrary thresholds, or established signatures. In this work, it is investigated if the angular power spectrum of the 3D structure factor can discriminate between morphologies. The 3D structure factor is computed on configurations of phase separating block copolymers generated by coarse-grained molecular dynamics simulations. The shell of structure factor values containing the primary peaks is isolated. This 2D field on a sphere is decomposed into spherical harmonic modes of even polynomial degree $\ell\le 12$, then further reduced to the rotationally invariant angular power spectrum. It is found that these few coefficients for low $\ell$ discriminate robustly between different morphologies. This analysis serves as an automatic tool for flagging novel structures, without a need to enumerate the plausible morphologies in advance.
\end{abstract}


\maketitle

\raggedbottom
\noindent In many applications of materials science, it is desirable to optimize a structural motif that affects material properties\cite{Lu2024-mq, Khadilkar2017-mo, Dong2023-vt, Beaucage2023-ae, Weeratunge2022}. However, structures of many material systems are dictated by thermodynamics rather than being directly chosen\cite{Matsen1996}. Therefore, multi-scale optimization processes tease apart how realistic design variables affect structure\cite{Li2024-ku, Robe2025, Weeratunge2023}, then how structure affects material properties\cite{Abdolahi2025, Shireen2022}. A central challenge is that “structure” is not generally simple to parameterize\cite{Lu2024-mq}, so intelligent exploration of the design space is limited to coarse categories\cite{Quah2025-wt, Ellis2025-xb}, or to preconceived set of structures\cite{Khadilkar2017-mo, Tsai2016-ym, Arora2021-qu}, or a specific target morphology (known in advance)\cite{Paradiso2016-hl, Dong2023-vt}. At its broadest scope, the motivation for exploring the method presented in this manuscript is to facilitate high throughput autonomous exploration of such morphological design spaces by constructing a metric that indicates discretely and robustly when a novel structure has been observed. A few studies introduce methods of highlighting truly novel structures, but they usually consider either free energy differences between structures\cite{Tsai2022-fs, Chen2023-kl, Dong2024-xa} or very high dimensional data\cite{Doerk2023-ij, Lu2024-mq}. The prior could obfuscate morphologies with similar energies (common in coexistence regimes), while the latter require significant effort to distinguish signal from noise. 

Block copolymers constitute a rich\cite{Park2003} design space of undiscovered\cite{Bates2012} microstructures. The classic system that still occasionally yields fruit\cite{Loo2005,Shi2013} is the diblock copolymer of species labeled A and B, characterized by a number of repeat units $N$, a volume fraction $f$, and an interaction strength $\chi$. If the product $\chi N$ is large enough, the energetic favourability of separating A from B overcomes the entropic favourability of mixing, and the system phase separates. Depending on the particular values of $f$,$\chi$, and $N$, the phase separated domains might form a variety of microstructural morphologies\cite{Matsen1996,Gavrilov2013}. The bulk material then has some characteristics of species A, some characteristics of B, and some characteristics that are due to the morphology. This means that block copolymer materials are straightforwardly tuned to modify material properties. Of course, block copolymers of more than two blocks, more than two species, nonlinear topology, and polydispersity are also manufactured, producing a wide variety of morphologies and material properties\cite{Tulsi2022,Liu2024,Dorfman2021}.

A central practice in the development of block copolymer materials is the construction of a phase diagram relating two or three design parameters to the resulting phase separated morphology\cite{Gavrilov2013,Dai2020,Zhao2021}. In some cases, the theoretical boundaries  between domains of different morphologies can be computed\cite{Matsen1996}, but empirical tracing of a phase diagram is traditionally a painstaking process of performing exhaustive measurements throughout the parameter space. Some phases occupy only thin slices of space between neighboring phases, so a dense grid of observations is needed. This creates a heavy combinatorial challenge for mapping out even two or three parameters. Modern adaptive sampling techniques\cite{Weeratunge2022,Ding2025,Weeratunge2025} can focus measurement effort on regions of parameter space near phase boundaries\cite{Dai2020,Zhao2021}, to more efficiently resolve the map. However, these techniques require a metric to quantitatively discriminate between different phases. Many methods exist to characterize structures and identify phases in a numerical model or an experimental measurement\cite{Akepati2025}, but they ubiquitously require an expert to construct some algorithm to detect a particular phase from a particular data set.

In the present work, we investigate if an analysis using angular power spectra derived from spherical harmonic decomposition can provide a parameter free, quantitative discriminator between microstructures with different morphologies.

Of course, this analysis is not limited to block copolymers or phase diagram sampling, but that application motivated the development of the method, and it is for that system which we will test the hypothesis that this analysis robustly discriminates between morphologies.

While it is advantageous to develop metrics that allow algorithms to navigate design spaces autonomously, it is also vital that these metrics are interpretable to humans. Contributing to understanding is after all the merit of simulation studies which are ultimately simplifications. Simulations and interpretable machine learning methods can work together\cite{Abdolahi2025, el, aplc} to optimize more efficiently\cite{Weeratunge2025} and discover trends\cite{Jayawardena2025, Amal2, Amal3, pop} if appropriate basis parameters can be identified. The angular power spectrum represents one such low-dimensional abstraction of diverse morphologies.

\section*{Results}
\subsection*{Angular Power Spectra}

\begin{figure*}[b]
    \includegraphics[width=.95\textwidth]{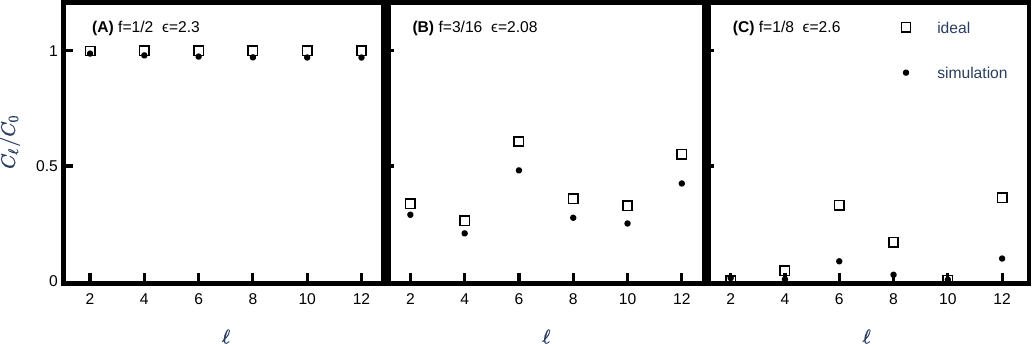}
    \caption{\textbf{Structural Angular Power Spectra (StrAPS).} Black circles represent postprocessing of morphologies found in MD simulations. Open squares represent idealized morphologies imposed by assigning particle types based on (A) lamellar, (B) hexagonal, and (C) BCC spherical phase fields.}
    \label{fig:straps}
\end{figure*}

Following the analysis detailed in the Methods Section, we carried out molecular dynamics (MD) simulations with coarse-grained bead-spring diblock copolymers of 16 beads each. Phase separation is induced by modifying the well depth $\epsilon$ of Lennard Jones interactions. Beads in blocks of different types interact with 1 kJ/mol, and beads in blocks of the same type interact with a system-specific value for $\epsilon$. We calculated the 3D structure factor for each simulation. We then radially averaged the 3D structure factor as usual to obtain the magnitude $k^*$ of wave vectors at which the primary peak occurs. We considered the 3D structure factor values for all wave vectors with magnitude near $k^*$ as samples of a structure factor field defined on a sphere. We performed a spherical harmonic  decomposition on those discrete points, and computed the angular power spectrum. We normalized the angular power spectrum for each simulation by the power at $\ell=0$. The resulting structural angular power spectra (StrAPS) for three different morphologies are presented as the black circles in Fig. \ref{fig:straps}. The first thing to note is that the StrAPS are qualitatively distinct for different volume fractions. Obviously, the StrAPS in Fig. \ref{fig:straps}A at $f=1/2$ with $\epsilon=2.3$ kJ/mol is nearly constant, while the StrAPS in B and C depend on $\ell$. Both B and C exhibit the highest structural angular power (StrAP) at $\ell=6$ and 12, but they are distinguished by disproportionately lower StrAP for $f=1/8$ (subfigure C) at $\ell=2$,4, and 10.

Without further analysis, the differences between StrAPS for different $f$ are empirical and the authors are unaware of any straightforward method to map general power spectra to morphologies. The aim of this analysis is to extract a compact feature vector that signals clearly when two systems contain different morphologies without manual labeling of every simulation, or even manual construction of standards. Once the presence of a (potentially novel) categorical distinction is known, manual morphology identification can be carried out for one or a few samples in each category.

In the case of these demonstrations in Fig. \ref{fig:straps}, the three different StrAPS are plausibly related to the well-established lamellar, hexagonal, and body centered cubic (BCC) spherical morphologies. These congruences are tested by generating an idealized configuration for each known morphology. We extracted particle positions from the end state of one of the aforementioned simulations, and re-assigned particle types according to idealized phase fields, regardless of original type or chain architecture. We then applied the same 3D structure factor, $k^*$ identification, spherical harmonic decomposition, and power spectrum analysis to each of these artificial configurations. The resulting StrAPS are included as empty square symbols in Fig. \ref{fig:straps}. The idealized phase fields were (A) lamellar, (B) hexagonal cylinders, and (C) BCC spheres.

Similar profiles could be predicted for any phase field for which a standard could be constructed. However, due to the discrete nature of morphological symmetries, these profiles are expected to be largely constant through regions of a phase diagram that map to the same morphology. Since this metric is compact and changes discretely as a function of design variables, it is a promising candidate for automatic clustering analysis during high-throughput exploration, or active learning algorithms that autonomously resolve phase diagram boundaries.

There remains the possibility that our simulation configurations are not actually in these expected morphologies, and their similar StrAPS are coincidental. To validate that the novel analysis presented here has not produced a “false equivalence”, we now consider the real space configurations and $k$ space structure factor.

\subsection*{Real Space Configurations}
 
\begin{figure*}
    \includegraphics[width=.31\textwidth]{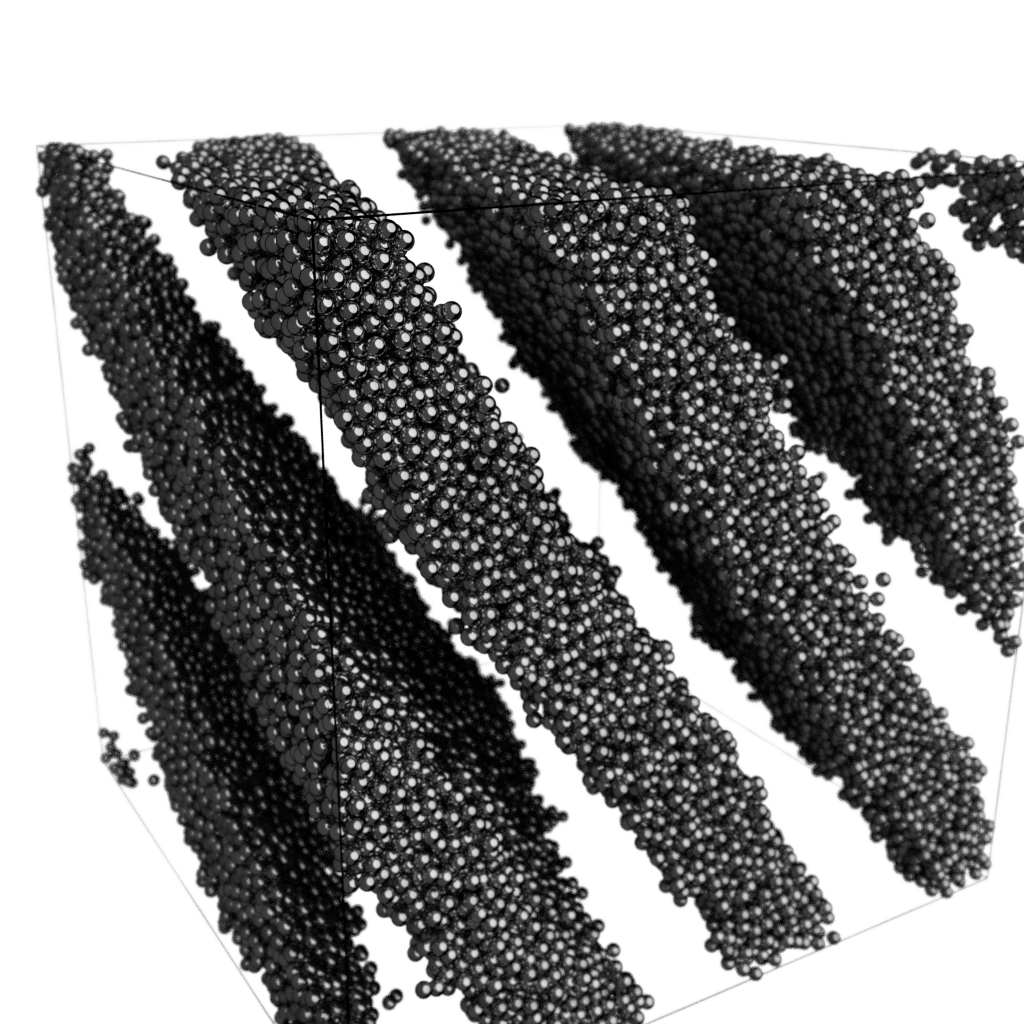}
    \includegraphics[width=.31\textwidth]{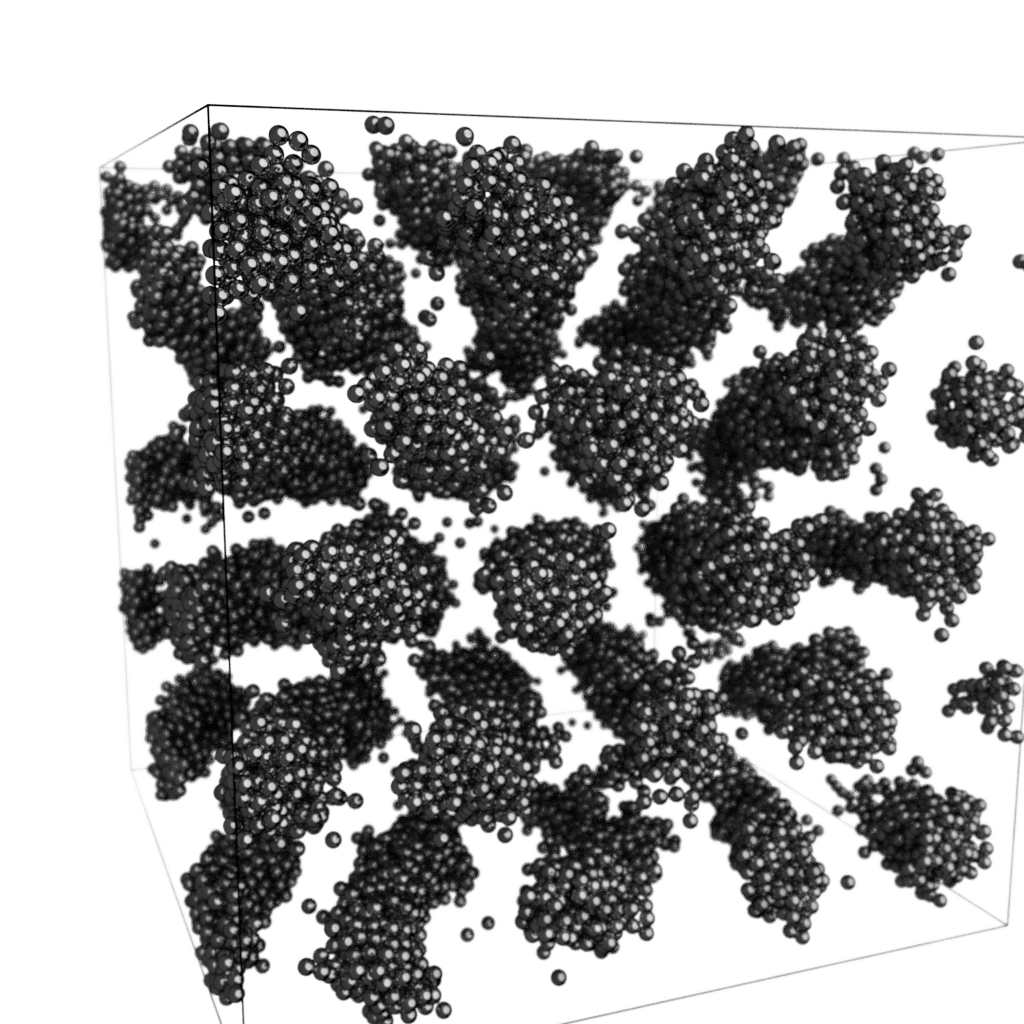}
    \includegraphics[width=.31\textwidth]{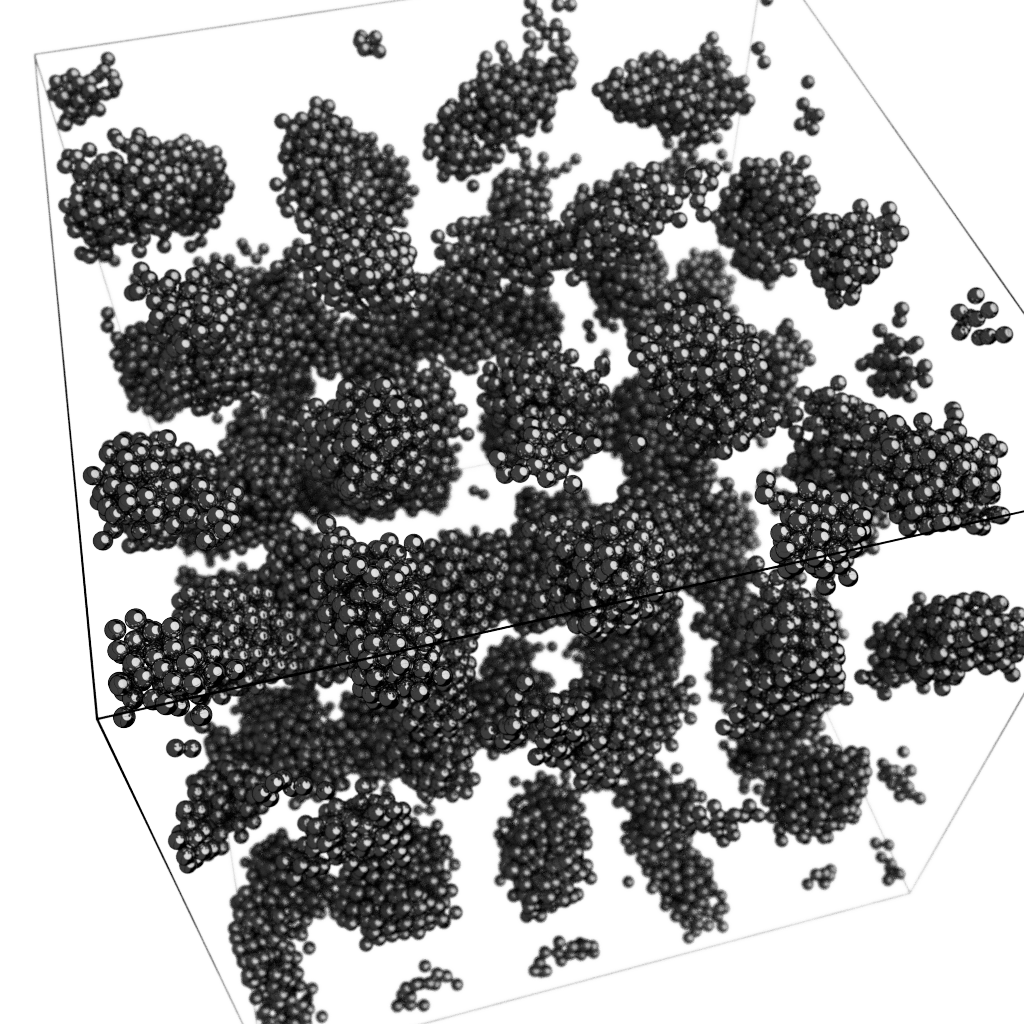}
    \caption{\textbf{Perspective renders.} Three different morphologies of phase separated microstructures. (Left) Lamellae at $f=1/2$. (Middle) cylinders at $f=3/16$. (Right) spheres at $f=1/8$.}
    \label{fig:perspective}
\end{figure*}

In Fig. \ref{fig:perspective} we show perspective renderings of the 3D configuration of particles in the minority block (type A). From left to right are shown the lamellar, cylindrical, and spherical microstructures, which are the same configurations used to calculate the StrAPS in Figs. \ref{fig:straps}A-C, respectively. The perspective rendering clearly shows that the first structure is formed of continuous layers, the second structure of linear cylinders, and the third structure of globular “spheres”.

\begin{figure*}
    \includegraphics[width=.95\textwidth]{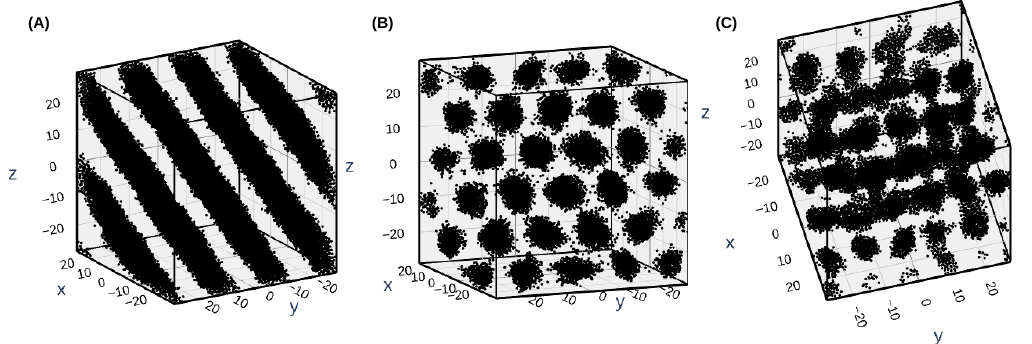}
    \caption{\textbf{Orthographic projections of particle centers.} These highlight the real space lattices of (A) Lamellar (B) Hexagonal and (C) BCC structures.}
    \label{fig:orthographic}
\end{figure*}

In Fig. \ref{fig:orthographic} we show orthographic projections of the particle centers, again just of the type A particles, to highlight the lattice structure of the phase separated domains. The layers in Fig. \ref{fig:orthographic}A are clearly flat and parallel, the cylinders in Fig. \ref{fig:orthographic}B are parallel and arranged hexagonally. The spheres in Fig. \ref{fig:orthographic}C are clearly arranged in a lattice that involves 90 degree angles.

Taken together, the views in Figs.~\ref{fig:perspective} and \ref{fig:orthographic} confirm that the StrAPS observed in Fig.~\ref{fig:straps} did arise from the expected morphologies. This does not prove that false equivalences are not possible. That is, there could potentially be two different morphologies with similar StrAPS, but at least in the examples we have considered, the correspondence is robust.

Notably, none of the morphologies presented here are aligned to the axes of the simulation cell. For lamellar and hexagonal structures this is plausible, but a rotated BCC (or gyroid\cite{Karatchentsev2010}) unit cell does not generally map to the cubic simulation cell. It is apparent from Fig.~\ref{fig:orthographic}C that the lattice is significantly disrupted. This is in part due to thermal fluctuations, which are also apparent in the lamellar and hexagonal configurations in 3A and B. However, in this case there turns out to be a secondary BCC arrangement of spheres. The square nature of the lattice highlighted in Fig.~\ref{fig:orthographic}C would be apparent from three different mutually orthogonal directions. The secondary lattice is apparent, but more distorted, from three mutually orthogonal directions as well, none of which are orthogonal to any of the clear perspectives on the other lattice. Rather than show these perspectives exhaustively, we will indicate the existence of these two lattices in the 3D structure factor data.

\subsection*{Structure Factor}

\begin{figure}
    \includegraphics[width=80mm]{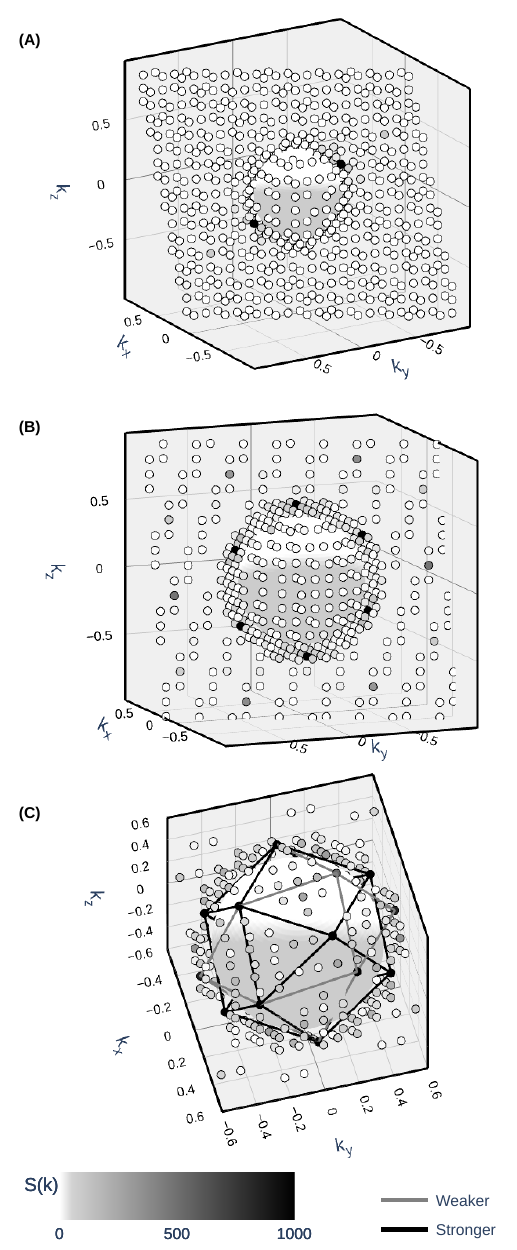}
    \caption{\textbf{Samples from the 3D Structure.} (A) lamellar, (B) Hexagonal, and (C) BCC morphologies. Values of $S(\mathbf{k})$ are presented for the shell of $\mathbf{k}$ vectors with magnitude $k^*$, as well as a plane in $k$ space. The darkness of each point indicates the value of $S(\mathbf{k})$. The gray and black wireframes in (C) indicate two competing BCC structures.}
    \label{fig:kspace3d}
\end{figure}

In Fig.~\ref{fig:kspace3d} we present a selection of the 3D structure factor measurements for our three different morphologies to build up an insight about how to interpret StrAPS. In each subfigure the camera is positioned (now in $k$ space) at the same orientation as in the corresponding panel of Fig.~\ref{fig:orthographic}, as this provides an informative view of the “peaks” in the 3D structure factor. For lamellar and hexagonal structures, all of the peaks lie in the plane normal to this camera’s position vector, so we have included all of the structure factor data in that plane. In 3D, the wave vectors making up the “peak” of the radially averaged structure factor at $k^*$ constitute a spherical shell of radius $k^*$. We have also included all of the points in this shell. Structure factor measurements also exist at all lattice sites in 3D $k$ space, but we have excluded all $\mathbf{k}$ outside of the specified plane and shell from our visualizations for clarity. The plotted plane and shell should be convincing that the 3D  $S(\mathbf{k})$ closely resembles delta function peaks at the ideal positions and a surrounding noise level several orders of magnitude lower than the primary peaks.  We have also included a white sphere slightly smaller than $k^*$ to visually suggest depth and to occlude points in the background.

In Fig. \ref{fig:kspace3d}A we see two clear peaks at wave vectors normal to the real space planes observed in Fig. \ref{fig:orthographic}A. The first satellite peaks are also apparent as gray points at 2 times the $\mathbf{k}$ vectors of the main peaks. In Fig. \ref{fig:kspace3d}B we see the six clear peaks in the $k^*$ shell. The orientation of each peak is normal to one of the planes formed by a row of cylinders in Fig. \ref{fig:orthographic}B. We can also see the dark gray points at a magnitude of $\sqrt{3}k^*$, rotated 30 degrees from the primary peaks, and four of lighter peaks at 2 times the main peaks. We note in passing that the second and third set of peaks form a hexagon in $k$ space, as would the fourth through sixth peaks and so on.

The 3D structure factor of the third system is more complex. A BCC lattice is represented by a cuboctahedron in $k$ space, with 12 vertices, 6 square faces and 8 triangular faces. The black lines in Fig. \ref{fig:kspace3d}C connect the 12 $\mathbf{k}$ points with the largest $S(\mathbf{k})$. In addition, there are six more $\mathbf{k}$ vectors with lower structure factor values still well above the background, which could be combined with 6 of the vectors from the other lattice to identify a new lattice rotated 60 degrees from the first. These points have been connected by gray lines in Fig. \ref{fig:kspace3d}C, forming a second cuboctahedron. The satellite peaks of this configuration are not confined to the plane normal to the camera, so are not visible in this representation. It bears repeating that the StrAPS does not require this degree of manual identification of patterns. The StrAPS were derived automatically from the simulation data, and the profile indicated a different kind of structure than the lamellar or hexagonal profiles. It could be said that the StrAPS “flagged” the presence of cuboctahedral ordering in spite of the fact that there were multiple overlapping patterns of peaks.

Fig.~\ref{fig:kspace3d} provides some intuition about why the StrAPS analysis discriminates between different structures. The angular power spectrum generally highlights variations characterized by a particular angle. The different structures contain different possible angles between peaks, so the StrAPS are distinct. The lamellar peaks are obviously separated by 180 degrees, and the hexagonal ones by 60 degrees. Depending on their relative positions, peaks in the BCC data might be separated by multiples of 60 or 90 degrees. The other obvious remark about Fig.~\ref{fig:kspace3d} is that the structure factor values in the $k^*$ shell other than the peaks are quite low. The effect of radial averaging on the signal-to-noise ratio could be counted as another reason the StrAPS are so robust relative to the averaged method.

\begin{figure*}
    \includegraphics[width=.95\textwidth]{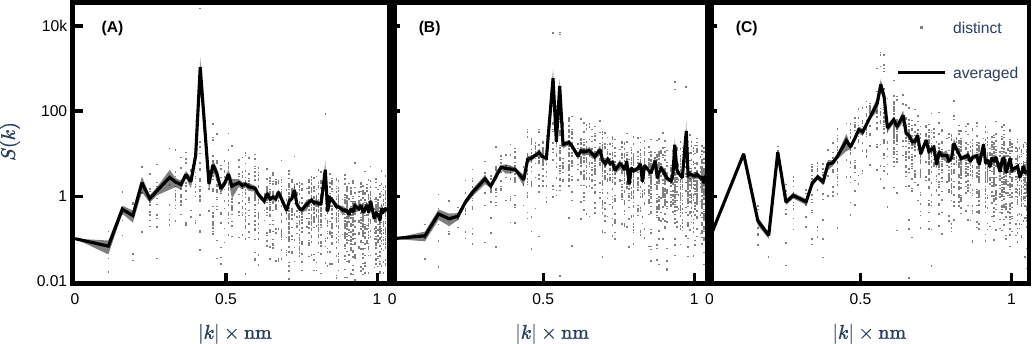}
    \caption{\textbf{Radially averaged structure factor.} Small gray points indicate values for individual 3D k vectors. The Black line indicates the average over vectors with the identical magnitude. A gray band around the averaged line represents the standard error of the mean for each k. Data are shown for the (A) lamellar, (B) Hexagonal, and (C) BCC systems.}
    \label{fig:kspaceradial}
\end{figure*}

Fig.~\ref{fig:kspaceradial} considers the effect of radial averaging to compare the effectiveness of StrAPS against the traditional method of microstructure characterization. Subfigures A, B, and C contain $S(k)$ data for the lamellar, hexagonal, and BCC systems, respectively. The small grey points indicate structure factor values for all points $k$ space, while the solid line shows the radial average. We note only exactly matching values of $k$ are averaged, and further binning would reduce the noise level but also shorten peaks. The $2k^*$ satellite peak is apparent for the lamellar data in Fig.~\ref{fig:kspaceradial}A, though is significantly reduced from the original height of the distinct peaks (gray points) due to the inclusion of many $\mathbf{k}$ vectors with exactly $2k^*$ magnitude but irrelevant angles. Similarly, the peaks in the Fig. \ref{fig:kspaceradial}B are significantly lower than they could be. Here the effect is worsened by the fact that a perfect hexagonal arrangement of $\mathbf{k}$ vectors is not allowed by a cubic simulation cell, so the peaks at various orientations with distorted magnitudes might end up in distinct $k$ bins, harming the signal-to-noise ratio.

In Fig. \ref{fig:kspaceradial}C, it is obvious that the signal strength of the peaks is weaker, even when separated by orientation. However, a few gray points may be observed reaching above the noise at $\sqrt{3}k^*$. This prominence becomes unconvincing after averaging, so the BCC character of this system would have been missed by any peak identification protocol applied to the radially averaged $S(k)$.

\section*{Discussion}

We do not endeavor here to calculate exact theoretical expectations for the StrAPS, as our motivation for developing the analysis was to discriminate between morphologies without the need to derive expected signatures. However, now that the analysis is shown to be informative, it could be fruitful to enumerate the signatures of known structures. It is also vital to probe systematically the robustness of these fingerprints. The two primary distortion mechanisms are thermal fluctuations of the phase boundaries and kinetic trapping in imperfect structures with grain boundaries or polydisperse domain sizes. If the effect of these imperfections is truly a uniform rescaling of the power spectrum, could it be predicted or mitigated? If the effect of distortions vary nonlinearly with the degree $\ell$,  what safeguards are needed to avoid identifying a malformed structure as novel?

A few limitations should also be noted. This method is only expected to be applied  directly to systems with long-range ordering, which produces the clear and ordered peaks in S(k). It is not clear if systems with many competing structures or a disordered morphology could exhibit any recognizable signal in the angular power spectrum. Another concern is that repeated stochastic simulations with different pseudo-random number generation seeds might produce multiple morphologies at the same nominal design variables. This is particularly concerning near phase boundaries, if there is genuine microstructure coexistence, or strong kinetic trapping. It could be considered a strength of this method that it would produce clearly distinct signals for each result, and near-zero power spectra for systems that either don’t separate or get trapped in disorder. However, it should be called out that this method does not resolve these fundamental problems around equilibration of phase separating systems. This analysis could also be used to optimize reliability by targeting regions of low variance of StrAPS across replicas.

The analysis presented here serves as a convenient way to distinguish between structural morphologies without the need for foreknowledge of the possible structures or their signatures, nor a method of extracting peaks from noisy scattering data. This work considers molecular dynamics simulations \cite{MD1, MD2, MD3, MD4, MD5, MD6} as a source of structural data that can be interrogated microscopically. Self-consistent field theory is also a common computational tool for investigating equilibrium of self-assembled microstructures of block copolymer systems. Even these coarser simulations have been accelerated\cite{Vigil2021-kz} or even surrogated\cite{Xuan2023-bv, Xuan2021-yt} by machine learning methods. These methods typically discriminate between morphologies by considering free energy\cite{Tsai2022-fs, Dong2024-xa}, traditional radially averaged structure factor peak detection, or mean curvature\cite{Chen2023-kl}. StrAPS is a complementary analysis that could support the detection of novel structures within any modeling paradigm for which 3D structural data is accessible.

Many algorithms have been developed to navigate the design space of soft matter systems. The proliferation of methods is due to the complexity of the "input" parameters, namely the full chemical specification of the material system. The "output" of these methods is typically either some encoding of a specific target, or a categorical label for a known structure. Continuous variable outputs have been limited to either extremely coarse (i.e. free energy) or extremely high dimensional (i.e. full structure factor). The StrAPS for a particular structure hold potential as a compact yet expressive vector of continuous features which could improve the ability of known exploration methods to navigate the output space as efficiently as they have been refined to handle the input space. It remains to be seen if phenomena such a phase coexistence and kinetically trapped samples can actually be leveraged as usable information to improve design space navigation with an appropriately chosen characterization method.

While in this work we have only shown StrAPS for the three most common classical morphologies, this proof of concept opens a path to a wide variety of applications, further tests, and extensions. It is particularly notable that the StrAPS for the BCC spherical morphology is clear enough to flag a point of interest, even though the disrupted structure is illegible in the traditional radially averaged analysis. We also note that we have calculated the StrAPS for a variety of other systems than the three presented here with different values of $f$ and $\epsilon$ and even multiple blocks per chain. The profiles from systems with different parameters but the same morphological class are generally constant. It is only in the case of deformed structures like that of the BCC system shown here that we observe a diminished profile. For the purpose of flagging potential novel morphologies, or for systematically refining the boundaries of phase diagrams, the method is viable. As discussed in the methods section, moving forward StrAPS can be used as a compact output feature vector for algorithms to facilitate exploration of phase diagrams, discovery of novel structures, and optimization of materials.

\section*{Methods}
\subsection*{Experimental Design}
To generate configuration data for phase separating systems, we performed coarse grained molecular dynamics (MD) simulations of block copolymers melts represented by bead spring chains using the many particle dynamics toolkit HOOMD-Blue\cite{ANDERSON2020109363}. Each simulation contained an ensemble of $N_c$ chains, each with $N_b$ beads. The fraction of beads $f$ is of type "A", and the remainder is of type "B". We have modeled diblock copolymers for this work, so the first $fN_b$ beads are type A.

Both A and B beads are represented as point particles interacting via Lennard Jones potentials with length scale $\sigma$=1~nm and Hookean springs with spring constant $k_s$=4~kJ/mol/nm$^2$. The Lennard Jones interaction energy $\epsilon_{AB}$ between particles of different type was fixed at 1~kJ/mol. The interactions between particles of the same type were set to $\epsilon_{AA}=\epsilon_{BB}=\epsilon$, which we used as a control parameter to explore the phase diagram of morphologies.

Chains were initialized with $\epsilon=\epsilon_{AB}$ in random walk configurations with a low density, energy minimized, and compressed over 10~ps to a density $N_b N_c/L^3=0.6$ using NVT integration. Here $L$ is the edge length of the simulation box. The temperature was fixed at T=293~K. The volume was then equilibrated for 10~ns using NPT integration with $P$=1~atm. The value of $\epsilon$ was then increased to its specified value, and a further 10~ns of NPT integration was performed. The equilibrated density depends mostly on $\epsilon$ and ranges from about 0.65~particles/nm$^3$ at $\epsilon$=2~kJ/mol to 0.75 at $\epsilon$=2.6. Finally, 100~ns (200~ns for $f$=1/8) of NVT integration was carried out at the equilibrated volume. We inspected simulation trajectories to ensure that morphologies were not continuing to evolve during the latter half of the final NVT run, and data presented in the Results Section represent the final configuration. Each simulation contained $N_c$=8000 chains. We carried out initialization and postprocessing in python.

\subsection*{Structure Factor}

The static structure factor $S(\mathbf{k})$ calculated from simulation data and experimentally measured small angle x-ray scattering (SAXS) intensity are the most common data sets from which morphology characterizations are drawn. The typical process is to radially average the $\mathbf{k}$ dependence, identify the $k$ values at which peaks in $S(k)$ are detected, then consider ratios of the peak $k$ values. Strong $S(\mathbf{k})$ signals in (3D) reciprocal space correspond to lattice planes in the real space morphology. A foundational body of work\cite{Matsen1996, Hamley1998-po} established the crystal structures of common morphologies and the resulting relationships between peaks in $S(k)$. Conveniently, for a given geometry, the "primary" peak at $k^*$ is the highest, and corresponds to the lattice constant of the morphology's unit cell via $a=2\pi/k^*$. Subsequent "satellite" peaks at higher values of $k$ correspond to reflections from various lattice planes, at predictable multiples of $k^*$ unique to each morphology.

We calculated the structure factor using

\begin{equation}
S(k)=\frac{1}{N_c N_b f}\left|\sum_{i=1}^{N_c N_b f} e^{i\mathbf{k\cdot r_i} } \right|^2,
\end{equation}

Where $\mathbf{r_i}$ is the position of the $i$th particle of type A. We carried out these structure factor calculations out for every $\mathbf{k}=k_0(i,j,k)$ where $k_0=2\pi/L$, $L$ is the edge length of the simulation box, and $i,j,k$ are any integer whose absolute value is less than $2/k_0$.

\subsection*{Spherical Harmonics and Power Spectrum}

The structure factor data for each simulation were radially averaged, and the wave vector magnitude $k^*$ of the maximum value of S(k) was identified. The criterion $k^*-k_0/2<k<k^*+k_0/2$ selects the $N_\mathrm{shell}$ vectors in the shell where $|\mathbf{k}|\approx k^*$. These points are taken as samples of the $S(\mathbf{k})$ field on that shell. The coefficients of the spherical harmonic decomposition of that field are computed as follows. The spherical coordinates ($r,\theta,\phi$) are computed for each vector $\mathbf{k_i}$ in the shell. The area $a_i$ of the shell represented by the point $\mathbf{k_i}$ is taken as the area of that point’s Voronoi cell on the sphere, calculated using the SphericalVoronoi function in the Python package SciPy. The coefficients of the spherical harmonic decomposition are then

\begin{equation}
c_{lm}=\sum_{i=1}^{N_\mathrm{shell}} {Y_\ell^m}^* (\theta_i,\phi_i)  S(\mathbf{k_i} ) a_i,
\end{equation}

Where ${Y_\ell^m}^*$ is the complex conjugate of the spherical harmonic function of degree $\ell$ and order $m$. 

\begin{equation}
{Y_\ell^m}(\theta,\phi)=A_\ell^me^{im\phi}P_\ell^m(\cos{\theta}).
\end{equation}

Here $A_\ell^m$ is a normalization constant and $P_\ell^m$ are the associated Legendre Polynomials

\begin{equation}
{P_\ell^m}(x)=\frac{(-1)^m}{2^\ell\ell!}\left(1-x^2\right)^{m/2}\frac{\mathrm{d}^{\ell+m}}{\mathrm{d}x^{\ell+m}}\left(x^2-1\right)^\ell.
\end{equation}

The power spectrum is then defined as

\begin{equation}
C_l=1/(2l+1) \sum_{m=-l}^l |c_{lm} |^2 .
\end{equation}

\subsection*{Improvement Opportunities}

Finally, we discuss some immediate directions for development of the StrAPS analysis. Some morphologies might not be fully characterized by the structure at a single length scale. One could easily expand the feature vector of the StrAPS to include higher degrees, or power spectra at multiple radii, though the first peak in $k$ space is highly important for many structures\cite{Tsai2022-fs}. One could also combine such feature vectors with the radially averaged $S(k)$, bearing in mind that piling on more features can make clustering more challenging. Here we have considered only the single-component structure factor, but more complex many-component materials might also require additional power spectra from each component. It is also possible to consider the full spherical harmonic decomposition instead of the power spectrum, but these coefficients are not rotationally invariant, so some justification would be needed for selecting a reference frame based on particular peaks of the structure factor. 3D X-ray scattering data is much more cumbersome to acquire than typical 2D patterns, but this method could be applied to such data where available. It may also be fruitful to consider a modified protocol to analyze 2D SAXS data\cite{Doerk2023-ij, Lu2024-mq, MD9}, though these typically contain averages over many grains.

Finally, the angular power spectrum could be computed for other fields than the structure factor. It could be informative to analyze local ordering in self-organizing systems that don't exhibit long-range periodicity. For instance one could compute the angular power spectrum for real space data at a particular length scale and average results from different locations since the power spectrum is rotationally invariant. An analysis of local structure fashioned after the method presented here could complement methods like CREASE\cite{Beltran-Villegas2019-df, Akepati2024-wu, Akepati2025-fg} for characterizing emergent structures. Such a method could also be applied to confined or patterned systems\cite{Shi2013, Liu2023-yw, sg1, sg2, sg3}.

\subsection*{Statistical Analysis}

As the main aim of this work is to compare the StrAPS to real space configurations and the convetional radially averaged structure factor, there has not been any averaging over time or simulation replicas. The only statistical analysis employed was to average structure factor observations with identical wave vector magnitudes in Fig.~\ref{fig:kspaceradial}. The number of data points in each bin varies according to the allowable wave vectors in a periodic simulation volume. The gray error band on those radially averaged data indicates the standard error of the data points at each magnitude.

\bibliographystyle{apsrev4-2}
\bibliography{StrAPSrefs}

\section*{Acknowledgments}

\subsection*{Funding:}

The authors acknowledge that they received no funding in support for this research.

\subsection*{Author contributions:}

Software: DMR

Visualization: DMR

Supervision: EH

Writing--original draft: DMR

Writing--review \& editing: DMR, EH

\section*{Competing interests:}

The authors affirm that we have no competing interests to declare.

\end{document}